# SDN-AAA: Towards the standard management of AAA infrastructures

Rafael Marin-Lopez, Oscar Canovas, Gabriel Lopez-Millan, Fernando Pereniguez-Garcia

*Abstract*—Software Defined Networking (SDN) is a widely deployed technology enabling the agile and flexible management of networks and services. This paradigm represents an appropriate candidate to address the dynamic and secure management of large and complex Authentication, Authorization and Accounting (AAA) infrastructures. In those infrastructures, there are several nodes which must exchange information securely to interconnect different realms. This article describes a novel SDN-based framework with a data model-driven approach following the standard YANG, named SDN-AAA, which can be used to dynamically manage routing and security configuration in AAA scenarios.

## I. INTRODUCTION

Software Defined Networking (SDN) [1] is a paradigm that has revolutionized the way networks are managed nowadays. SDNs are characterized by the separation of the *control plane*, the part of the network node which takes decisions, and the *data plane*, which is in charge of forwarding packets as per the instructions of the control plane. The control plane is now centralized in an entity named *SDN controller* that can configure the data plane of multiple network nodes through an interface (*southbound interface*). This paradigm is being used in many different applications to manage security or IP forwarding processes. For example, in the IP forwarding application, the controller (control plane) provides rules to the forwarding tables of SDN-capable network nodes (data plane) so that they only forward the IP packets based on the information provided by the controller [2].

The Authentication, Authorization and Accounting (AAA) infrastructures [3] have been used for decades for controlling the access to network resources (e.g. network connectivity) by verifying the identity (authentication) of the supplicant entity that tries to get access to the network resource; for determining which privileges are assigned to the supplicant (authorization); and, for registering information about the usage of resources (accounting). AAA infrastructures are formed by a set of interconnected AAA nodes. Each AAA node is expected to (1) establish adjacency relationships with other AAA nodes (known as *peers*) by setting up a secure communication channel to protect the exchanged AAA information, and (2) to forward the messages that contain the user's authentication and authorization information to the next hop in the path, so that it eventually reaches the AAA node that can verify that information.

Nowadays, AAA infrastructures are an essential part of the cellular networks (4G and 5G) [4], controlling the access of thousands of devices and, moreover, it is the central infrastructure in the international Wi-Fi internet access roaming service named *eduroam* [5]. Hence, the number of users that this kind of infrastructures must control has increased substantially. This makes their maintainability difficult and manual configuration must be avoided at all cost, specially when the AAA infrastructure is complex (i.e., there exists a high number of interconnected domains, users and connections among AAA nodes). Furthermore, considering the dynamism of current networks, where AAA nodes can be virtualized and deployed on demand [4], and the increasing number of SDN-based scenarios (e.g. the Internet-of-Things [6]), a mechanism to automate the routing management of AAA infrastructures according to the SDN paradigm becomes even more relevant. Following the SDN principles, this paper proposes a solution called SDN-AAA, which uses well-known standards such as NETCONF and YANG to dynamically manage AAA infrastructures. More specifically, SDN-AAA provides:

- *Bootstrapping of security associations between AAA nodes on demand*. SDN-AAA allows centralizing the configuration of different types of credentials (certificates, raw public keys, shared keys, etc.) to protect the AAA traffic between any two adjacent AAA nodes.
- *Dynamic configuration of AAA routing information*. Routing management enables dynamic forwarding of AAA traffic (flows) between different AAA nodes. SDN-AAA can adapt quickly the routing information when a new AAA route needs to be set up, or when a AAA node is not operative or congested and the AAA traffic needs to be forwarded through a different AAA node.

Controlling the traffic flows of authenticated and authorized network nodes with SDN have received great attention [7] in the related literature. Most of the existing works [6], [8], [9] rely on the usage of AAA infrastructures not only to authenticate and authorize end users but also to instruct the controller in the domain to send rules to the network nodes to enforce policies and authorization information about traffic flows crossing the network of authenticated end users. However, these works do not consider the usage of SDN to manage the AAA infrastructure itself allowing the routing of AAA information to the proper AAA server. Similarly, the

Gabriel Lopez-Millan and Rafael Marin-Lopez are with the Department of Information and Communications Engineering , University of Murcia, Spain.
Oscar Canovas is with the Department of Computer Engineering, University of Murcia, Spain.
Fernando Pereniguez-Garcia is with the Department of Engineering and Applied Technologies, University Defense Center, Politechnic University of Cartagena, Spain.
Manuscript received XX

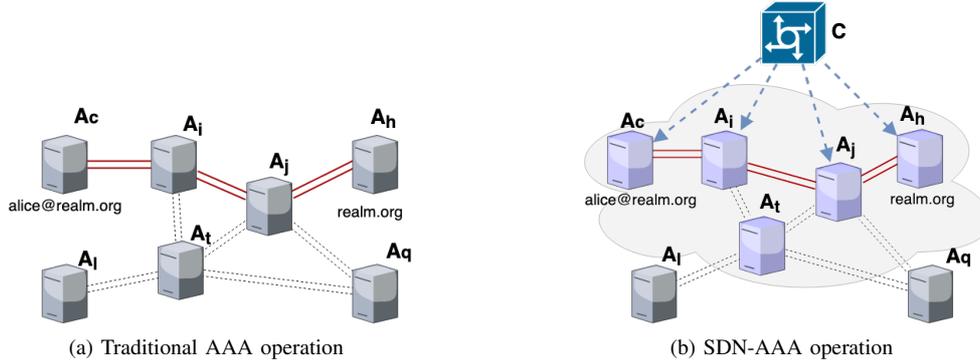

Fig. 1: Two different alternatives of AAA operation

applicability of SDN to manage IP routing is well known and, in fact, constitutes one of the main applicability areas of SDN. We can find standards developing YANG data models for the configuration of well-known routing protocols like RIP or OSPF, but not for AAA infrastructures. Academic research works in this area focus on the definition of novel SDN-assisted routing protocols for concrete scenarios like vehicular or Internet-of-Things networks [6], [10]. For example, [10] focuses on the definition of a secure routing solution for IoT (the whole packet is encrypted with a shared secret) assisted by a SDN controller. In this sense, as far as authors' knowledge, our proposal constitutes the first attempt to provide mechanisms for the management of AAA routing under the SDN paradigm. Moreover, we also define and detail an initial design and a data model based on the standard YANG.

The remainder of this article is organized as follows. Section II describes the AAA routing process and some related issues. Section III describes the SDN-AAA framework in detail. Section IV explains the designed YANG data model. Section V exposes some security considerations. Finally, Section VI concludes the article and provides some future directions.

## II. ISSUES RELATED TO AAA MANAGEMENT

AAA infrastructures are formed by a set of AAA nodes. Each node maintains an adjacency relationship with other AAA nodes (peers) by means of a AAA protocol. The most widely deployed AAA protocols are RADIUS, generally used in WiFi deployments like *eduroam*, and Diameter, typically used in 4G/5G networks. AAA messages must be protected by using well-known security protocols such as (D)TLS or IPsec between adjacent nodes. Each AAA node may play a different role:

- **AAA client** is a AAA node that implements the client part of a AAA protocol and it is usually an entry point to the AAA infrastructure.
- **AAA server** handles authentication, authorization, and accounting requests for the users of a particular realm. Usually, it is called *visited server* when the user requesting access does not belong to its realm, and *home server* when we are referring to the AAA server able to validate the credentials provided by the users of this realm.
- **AAA agent** can be a *Relay*, which forwards AAA requests to another adjacent node (known as next-hop in routing terminology) depending on the user's realm; *Proxy*, which also routes AAA messages but, unlike relay agents, it may modify some attributes contained in the messages to enforce policies; *Translator*, which translates from one AAA protocol to another (e.g. RADIUS to Diameter or vice versa); and, finally, *Redirect*, which provides routing information to any requesting peer.

In a typical AAA process, a user who wants to access a network resource (e.g., wireless connectivity) must provide their identity and credentials to be authenticated and authorized (see Figure 1). The identity has the format of a network access identifier (NAI), that is, a user name and a realm where the user is registered to. For example, *alice@realm.org*, where *alice* is the user name and *realm.org* is the realm to which Alice belongs. This information is taken by the device controlling the access to the network resource (e.g. a Wi-Fi access point), which typically implements the AAA client. Based on the realm contained in the identity, the AAA infrastructure routes the AAA protocol messages through different AAA nodes from the AAA client ($A_c$) to the *Alice*'s home server ($A_h$), which is able to verify Alice's credentials.

To achieve this, each AAA node must know his adjacent nodes' information (IP address or host name, ports,...) and establish a security association with them when necessary during the routing process. In this example (Figure 1a), adjacent nodes of $A_i$ are $A_c$, $A_t$ and $A_j$. For this reason, every node maintains the so-called *peer table* to store the information of the adjacent nodes and to establish these security channels. Additionally, each node also maintains a *routing table* used to determine the next adjacent node (next hop) where to forward the messages. In this case, for example, the $A_i$'s routing table points towards $A_j$, as the next hop to reach the destination realm (*realm.org*). In the example, the path between the AAA client ($A_c$) and the home server ($A_h$) involves $A_i$ and $A_j$. Thus, three security channels are established: between $A_c$ and $A_i$; $A_i$ and $A_j$; and, finally, $A_j$ and $A_h$.

Typically, both the peer table and the routing table of different nodes are manually configured by the administrator. Therefore, the AAA infrastructure may not be resilient to the failure of some of the nodes in the path because the routing tables are not dynamically updated but statically configured. For example, RADIUS traditionally builds the routes from a configuration file. If one of the peers is down, there is no standard mechanism to dynamically look for new nodes serving as next hop towards a concrete realm and to change these routes.

On the contrary, Diameter provides two mechanisms. Firstly, to discover a new peer using information from the DNS (e.g. by using NAPTR records), with the establishment of (D)TLS security channel. Nevertheless, the current standard is limited to the usage of digital certificates to establish security associations between AAA nodes and recognizes the need of a Diameter server Certificate Authority (CA), which is not being deployed yet in the case of Diameter. It is worth noting that certain scenarios have adopted this kind of solutions when using RADIUS as well [5]. Secondly, to query a Diameter Redirect agent, which holds routing information. Unfortunately, the configuration format of this information for Diameter Redirect agents is not defined.

In summary, the administration of realms may become a cumbersome task, with numerous ad-hoc entries and manual configuration of AAA servers. Additionally, in hierarchical topologies, there are performance issues derived from the routing of AAA messages that concentrates large amounts of traffic in the proxies situated in the upper levels (i.e., there is no end-to-end relationship between the AAA server in one domain and any other AAA server at the same or lower level of the hierarchy). Finally, the more nodes are needed, the higher packet loss probability, taking into account the possible congestion that certain AAA nodes in the hierarchy may suffer [5].

In our proposal, Figure 1b, the SDN controller takes the responsibility of establishing the security channels between adjacent AAA nodes (i.e. filling the peer table) and to send the routing information (i.e. populating the routing table). In other words, the controller ($C$) is able to instruct the AAA nodes under its control about how to reach any particular realm in the AAA infrastructure. Following the SDN paradigm, the controller can modify dynamically the topology of the AAA infrastructure redefining the peers of a specific node and establishing the pertinent security channels. Moreover, it can modify the routes in the AAA infrastructure when necessary. In this way, it can overcome the failure of nodes, unexpected traffic growth, the support for new realms, etc. Moreover, this general operation is independent of the AAA protocol.

## III. SDN-AAA: SDN-BASED AAA MANAGEMENT

### A. General architecture

The proposed framework is compliant with the general SDN architecture and, consequently, composed by two entities: the controller, which represents the control plane, and the AAA nodes, which constitute the data plane.

Figure 2 shows the general architecture of SDN-AAA with the typical layering in the SDN paradigm (application plane, control plane and data plane) mapped to the application of AAA management. In the *application plane*, the administrator can provide information about the AAA management policies through the *northbound interface*. These policies describe how to manage the AAA data plane (domain routing policies, security policies, etc.). For example, an administrator could specify something like: "route domains *.org to aaa-node.realm.org and provide TLS security associations between AAA nodes". The software module executed in the application plane allows defining those policies in a high level description language. Currently, there is a clear tendency towards implementing the northbound interface using RESTful communication models and YANG as data modelling language to define the structure of exchanged data.

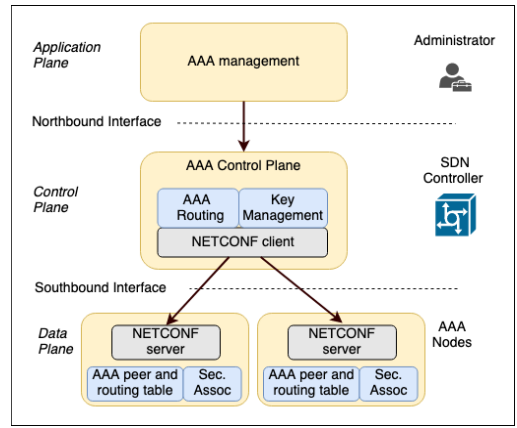

Fig. 2: General architecture of SDN-AAA

These general policies are to be implemented by the *control plane*, where the controller has a AAA Routing software module with information about the adjacent AAA nodes (peers) in the topology (peer table) and the routing information, which will be useful to configure accurately the AAA nodes' routing table. Moreover, it implements a Key Management software module which permits to configure security associations (IPsec or TLS) between any two adjacent AAA nodes.

At the *data plane*, the AAA node has the peer table, with the adjacent nodes (peers)' information, such as IP address or credentials to establish a security channel to protect the AAA traffic between the node and the peers; and the routing table, with information about the next AAA node to reach a particular realm. The detailed set of data to configure is described in Section IV. These nodes are configured by the controller through the *southbound interface*.

The main function of this interface is to provide a secure communication channel between the controller and the data plane nodes. In the context of SDN-AAA, this channel is used for several tasks: to apply peer information, security configuration and routing information to AAA nodes, to recover AAA state data from the AAA nodes, and to allow the AAA nodes to send notifications to the controller. Configuration, state and notification data are AAA protocol independent.

SDN-AAA opts for implementing the southbound interface using the NETCONF management protocol together with YANG, a de facto standard as data modeling language. Despite there exists other candidate southbound protocols like

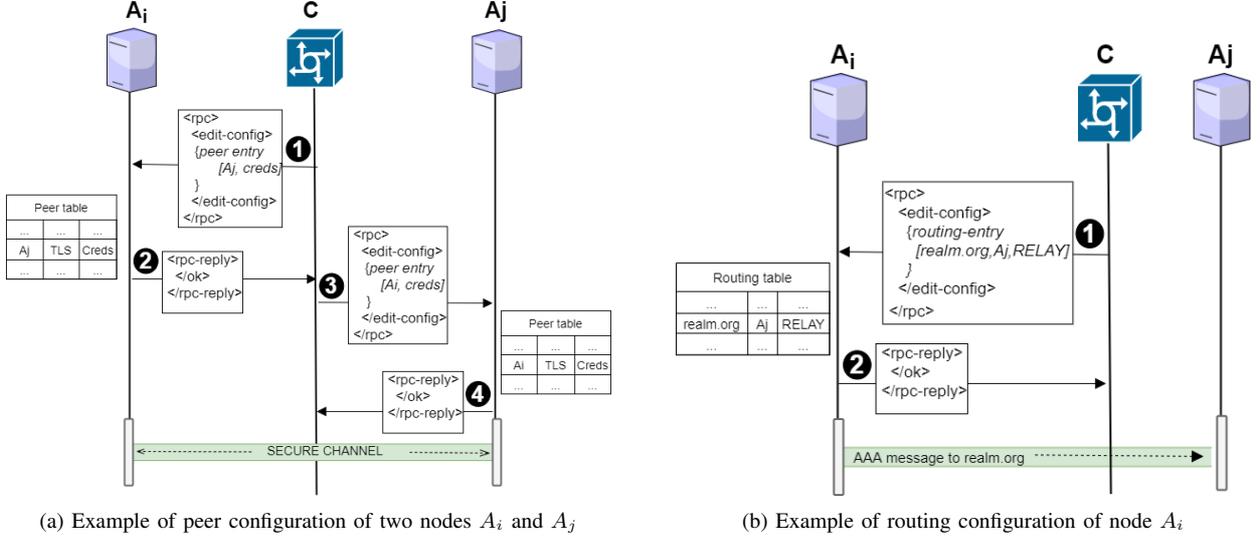

(a) Example of peer configuration of two nodes $A_i$ and $A_j$

(b) Example of routing configuration of node $A_i$

Fig. 3: SDN-AAA operation

SNMP or OpenFlow, we have selected NETCONF because, along with YANG modeling, provides much more flexibility than SNMP to represent AAA configuration with a generic representation. Beside, NETCONF allows using a myriad of standard and proprietary YANG data models, in contrast with Openflow, which was specifically designed for SDN-based layer 2 network traffic forwarding.

Specifically, NETCONF provides a secure communication channel between the controller and the AAA nodes, based on SSH (by default) or TLS. NETCONF operations (*edit-config*, *get-config*, *notifications*, etc.) are used to configure and monitor the managed AAA nodes.

### B. General operation

We assume that the AAA nodes are already deployed in the network and registered. This step is assumed by any SDN-based application and, therefore, out-of-scope of this article. The controller is in charge of configuring the nodes to route the AAA information properly. In order to achieve this, the controller implements two basic operations:

1) Configure the peer table of AAA nodes in order to specify what are the adjacent nodes and type of security to protect the information between them.
2) Configure the routing table of AAA nodes to specify the next AAA node (next hop) required to reach a certain realm.

Let us consider the example where the controller $C$ needs to establish a route in $A_i$ with $A_j$ as next hop to reach a realm *realm.org*. As such, $A_i$ and $A_j$ are considered adjacent nodes. Consequently, the controller must configure first the peer table of $A_i$ by specifying that $A_j$ is the adjacent node and vice versa. Following the procedure depicted in Figure 3a, the controller firstly sends *edit-config* message to $A_i$ (*step 1*) with a peer entry: the $A_j$'s information (e.g. IP address/host identity, ports, etc.), and credentials to establish a security channel with $A_j$ (e.g. TLS credentials). When the controller receives the $OK$ message (*step 2*), $A_i$ has the peer table configured. Similarly, it sends to $A_j$ a peer entry with $A_i$'s information as the adjacent node and the credentials to establish the security channel (*step 3*). When the controller receives the $OK$ message (*step 4*), both nodes are configured and can establish a secure channel. From this point, both are considered adjacent nodes so the controller can send now routing information. It is worth noting that the controller could alternatively send the messages in *step 1* and *step 3* in parallel to reduce the overall total time. If some of the operations described fails, the controller performs a rollback operation in the node that was properly configured to remove the entry in the peer table.

Once the adjacency relationship has been established between $A_i$ and $A_j$, the controller follows the process shown in Figure 3b to provide $A_i$ with the necessary routing information. The process starts when the controller sends to $A_i$ the routing entry (*step 1*) that indicates that $A_j$ is the next hop to reach the domain *realm.org*. It also configures the expected action in the node (e.g. RELAY), which implies the mere forwarding of the message. For the next hop, between $A_j$ and any other AAA node, a similar process would be also required.

It is worth noting that if the controller already knows the routing information applicable to $A_i$, it can optimize the aforementioned exchange in Figure 3a by sending not only the peer entry information but also the routing information in the same *edit-config* message in *step 1*. Finally, if two nodes were already configured as peers, the controller only needs to send the routing information.

With these basic operations, the controller can implement two modes of operation: proactive and reactive. In the *proactive mode*, the controller installs the required information in the peer table and the routing table of the AAA nodes *before* the AAA messages need to be routed. In this manner, when a AAA message arrives, all the information required is already in place. Conversely, in the *reactive mode*, the controller waits for a notification from the node to install the information required (peers and routing) in the AAA node. This notification is sent *after* the AAA node verifies that it does not have the information about how to route a AAA message to a specific realm.

## IV. A YANG DATA MODEL FOR SDN-AAA

As we already mentioned, SDN-AAA uses NETCONF as southbound protocol and YANG as language to model configuration data exchanged between the controller and the AAA nodes. A YANG data model follows a tree structure of different types of data nodes. For example, the type *container* is used for nodes residing in intermediate levels of the tree that contain other nodes, thus grouping data logically. Conversely, the type *leaf* is used for end nodes containing simple data (e.g. a numeric value). We can find previous efforts of standardization bodies and vendors to provide YANG models for AAA equipment but mainly focused on specific AAA protocols or implementations.

As a central part of our contribution, we have provided a YANG-based data model for SDN-AAA which defines a generic interface to program routing and security features on AAA nodes, regardless the specific AAA protocol and implementation. Our YANG data model has been validated in a software prototype and has been derived from the specifications of RADIUS and Diameter standards, besides analyzing some open source implementations, such as FreeRADIUS, RadSecProxy or freeDiameter, in order to ensure that there is a map between the proposed model and the configuration options of real software packages. Since our current model provides the definition of security channels for TLS-based applications, we make use of the YANG data model for configuring TLS connections [11].

As Figure 4 shows, the YANG data model consists of several parts related to different configuration containers. The two central elements that we describe in detail are the peer and the routing entries.

The $peers$ container consists of a list of $peer-entry$, each one referring to an adjacent AAA node that can act as source or destination of AAA messages. The main information elements for a peer entry are the identity of the peer, the expiration time for the entry, the credentials required to protect the AAA messages exchanged with the peer, and the transport method. In relation to this latter element, common values are RADIUS over UDP, RADIUS over TLS, or Diameter over TLS/TCP. On the one hand, when TLS is used, we provide in our data model several information elements to establish the TLS-based security channel in a different container named $tls$. It consists of a list of TLS profiles that can be used by the node when it is required to establish TLS channels with other nodes. Mainly, each profile contains information about the identity to be used, including cryptographic material or references to stores where the credentials can be found, and potential requirements to authenticate other peers based on certificates, keys or specific attributes. On the other hand, if TCP or UDP are used then it is required to use other SDN-based approaches for dynamic establishment of security associations, like the standardized YANG data model for IPsec flow protection [12].

The $routing$ container is composed by a list of entries ($realms-entry$) describing how the messages related to a specific realm must be routed. Some of the main information elements for each entry are the destination realm where the messages should be routed, the identifier of the peer entry considered as the next hop for the realm, an action to perform with the AAA message, attribute management rules defined in a separate $attributes$ container to modify incoming or outgoing messages, and an expiration time for the entry. In relation to the action to perform, given a destination realm, the AAA node can take different actions with the AAA message, that is, to process it locally (LOCAL), just to forward the message to the next peer (RELAY), to apply some modification to the AAA message before forwarding it to the next hop (PROXY), or to redirect the AAA message to a AAA redirect agent to obtain information about the next hop (REDIRECT).

Additionally, the module defines notifications to allow the node to inform the controller about the need to acquire routing information for unknown realms, as it was previously explained for the reactive mode, or when a peer or routing entry expires. Notifications contain a list of attributes. For example, an attribute containing the realm is required for the acquire notification since it will be used by the controller to determine the routing data.

## V. SECURITY CONSIDERATIONS

A formal security analysis of any SDN-based solution is complex, mainly because the number of interfaces, protocols and entities involved. Previous works, such as [13], have well analyzed the security aspects related to SDN. We summarize here some general considerations related to SDN-AAA.

### A. Access Control

An attacker could be able to carry out different types of attacks over the communication path between the controllers and the nodes, such as eavesdropping, traffic analysis, spoofing, insertion, modification, deletion, delay, replay or impersonate any of the entities. This means that the attacker might read messages on the network and remove, change, or inject forged messages. Nevertheless, to prevent this kind of attacks, it is required to enforce strong access control mechanisms (i.e. authentication of the entities), to establish authenticated, and well-protected communication channels with integrity and confidentiality between the controller and nodes (southbound interface). The proposed framework makes use of NETCONF, which mandates the use of a secure transport protocol like SSH (by default) or TLS, which provides confidentiality, integrity and mutual authentication.

Additionally, it is necessary to restrict access to certain YANG elements by means of the NETCONF Access Control Model (NACM) [14]. This sensitive information ranges from AAA routing configuration parameters to the cryptographic material required by the AAA secure communication channels.

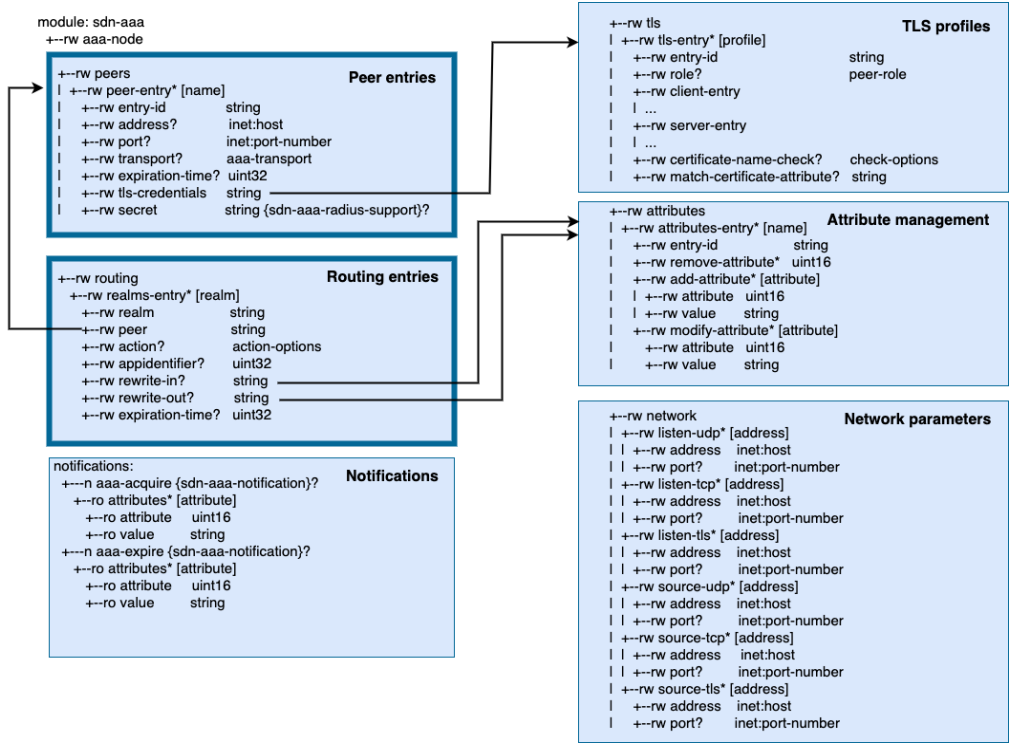

Fig. 4: Tree of the proposed SDN-AAA YANG model

*B. Confidentiality and integrity*

There are two main processes in SDN-AAA that, in case of attack, can affect confidentiality and integrity of the AAA protocol. The first one is the generation of the cryptographic material required by the security protocol used between AAA nodes. For example, shared secrets for basic RADIUS/UDP, pre-shared keys or public/private RSA keys for RADIUS/TLS or DIAMETER over TLS. The second one is the key material distribution process itself, when the controller sends this information to the nodes.

The controller is in charge of generating the cryptographic material for encryption and integrity protection depending on the security requirements of the network security protocol used in the AAA data plane. Therefore, the controller is potentially capable of decrypting any communication between any pair of nodes, affecting the confidentiality and integrity of the communications. Some requirements in order to avoid or reduce the possibility of an attacker to access this key material are the following: the node must not allow the reading of private cryptographic material (i.e., write-only operations); if PSK authentication is used the controller must generate randomly the PSK and remove it immediately just after being distributed in order to reduce the impact if an attacker has been able to compromise the controller; if RSA keys are used, and the controller generates the key pairs, it must remove the associated private keys immediately after distributing them to the nodes for the same reasons as above; the ciphersuites used for either NETCONF over SSH or TLS must generate key material that allows protecting the information exchanged between the controller and the AAA nodes.

*C. Availability*

The controller represents a single point of failure and a target for attacks. In case of a controller failure, the AAA network security channels between nodes will not be monitored by the controller. The consequences is that if the communication channel fails, or the security association expires, the controller will not be able to manage the data plane and the AAA network will be disrupted. In SDN networks, solutions for high availability and load balancing for the controller (i.e. having several instances of the controller) are deployed to alleviate these problems [15].

Alternatively, if the controller is attacked, it can severely affect the operation of the network, provoking its complete disruption. In terms of the AAA process, it may start a denial of service attack over the nodes installing fake AAA routes. Moreover, if nodes are attacked, the attacker can force the controller to start routing requests by sending routing notifications even though they are not necessary. In this case, general solutions for detecting and mitigating the effect of DDoS can be applied.

VI. CONCLUSIONS

In this article, we have defined the essential pillars for a flexible and dynamic management of the AAA infrastructure operation based on the SDN paradigm. We have called our proposal SDN-AAA, which is based on solid standards such as NETCONF and YANG. In particular, SDN-AAA allows a dynamic AAA routing within the infrastructure. To achieve this, the SDN controller is able: 1) to establish the relationships of adjacency between AAA nodes (peers) providing all the parameters to establish security channel between them, and 2) to provide the routing information, based on the user's realm, to reach the proper AAA server in charge of the user's authentication and authorization.

Besides explaining the architecture and the operation of our proposal, we have defined and tested an initial YANG data model that incorporates the parameters required to configure dynamically the nodes of the AAA infrastructure. We have modelled the peer table, with the security parameters required to protect the communication between them, and the routing table. This is just the first step to have a complete AAA management solution with SDN. As a statement of direction, we are working on an implementation that is being tested in virtualized environments to obtain performance results.

**Rafael Marin-Lopez** Rafael Marin-Lopez is a full-time assistant lecturer in the Department Information and Communications Engineering at the University of Murcia (Spain). Additionally, he is collaborating actively in IETF, especially in I2NSF and ACE Working Groups. He is co-author of RFC 9061, RFC 7499 and other standards. His main research interests include network access authentication, key distribution and security in different type of networks such as SDN, IoT and mobile networks.

**Oscar Canovas** Óscar Cánovas was born in Barcelona (Spain), in 1975. He received his Ph.D. degree in Computer Science from the University of Murcia, Spain, in 2003. From 1999 to 2009 he was an Assistant Professor with the University of Murcia. Since 2009 he has been an Associate Professor at the same university. His research interests include information security, access control, ubiquitous computing and indoor positioning.

**Gabriel Lopez-Millan** Gabriel Lopez-Millan is a full-time assistant professor in the Department of Information and Communications Engineering of the University of Murcia. He has collaborated actively in IETF WG such as ABFAB and I2NSF. He is co-author of RFCs 9061 and 7499. His research interests include network security, SDN/NFV, PKI, identity management, authentication and authorization. He received his PhD in computer science from the University of Murcia.

**Fernando Pereniguez-Garcia** Fernando Pereniguez-Garcia received his Ph.D. degree in Computer Science in 2011, from University of Murcia (Spain). Currently, he is Assistant Professor in the Department of Engineering and Applied Technologies at University Centre of Defence, Politechnic University of Cartagena. He is co-author of RFCs 9061 and 7499. His research interests include network security and privacy in SDN and wireless mobile networks.